\documentstyle[]{article}
\newlength{\aivwidth}   
\newlength{\tmpwidth}   

\title{ Statistics of skyrmions in quantum Hall systems }
\author{Jacek Dziarmaga  \\
        Department of Mathematical Sciences,\\
        University of Durham, South Road, Durham, DH1 3LE,\\ 
        United Kingdom.\\
        E-mail address: J.P.Dziarmaga@durham.ac.uk 
        \thanks{On leave of absence from Institute of Physics,
        Jagiellonian University, Krak\'ow, Poland} }
\date{December 11, 1995}
\begin{document}
\maketitle
   \begin{abstract}
   We analyze statistical interactions of skyrmions in the quantum
Hall system near a critical filling fraction in the framework
of the Ginzburg-Landau model. The phase picked up by the wave-function 
during an exchange of two skyrmions close to $\nu=1/(2n+1)$ is 
$\pi[S+1/2(2n+1)]$, where $S$ is the skyrmion's spin. In the same setting 
an exchange of two fully polarized vortices gives rise to the phase  
$\pi/(2n+1)$. Skyrmions with odd and even numbers of reversed 
spins have different quantum statistics. Condensation of skyrmions with an 
even number of reversed spins leads to filling fractions with odd denominators,
while condensation of those with an odd number of reversed spins
gives rise to filling fractions with even denominators.
   \end{abstract}
\vspace*{1cm}
DTP-95/79\\
\vspace*{1cm}

Recently skyrmion excitations near the filling factor $\nu=1$
have drawn remarkable attention \cite{sondhi,fertig,moon}. There are two 
kinds of topological excitations in single-layer quantum Hall systems.
When the system is fully polarized, the relevant charged quasiparticles
are topological vortices. At $\nu=1/(2n+1)$ the charge of such a vortex is 
$e/(2n+1)$ and the spin is $1/2(2n+1)$. It has been suggested \cite{sondhi} 
that for a weak Zeeman coupling the lowest energy quasiparticle is a skyrmion 
or a slowly varying topological texture. Its electric charge is still 
$e/(2n+1)$ but its total spin $S$ can be substantially larger. Such large 
spin but moderate charge excitations can explain the observed depolarization 
as the filling factor slightly deviates from $\nu=1/(2n+1)$ \cite{knight}.

  It has been proposed to describe skyrmion excitations in terms of an
effective ferromagnetic model \cite{sondhi}. The model explains the 
dependence of the skyrmion size on relative strenght of Coulomb and 
Zeeman interaction. However, it does not predict correctly  skyrmion 
energies. The energy of skyrmion and antiskyrmion seem to be the 
same. Microscopic calculations \cite{fertig} show that it is not the 
case. The effective model has been derived \cite{sondhi,stone} from 
the Ginzburg-Landau model of the quantum Hall system \cite{zhk,kl}. In the  
derivation, among other things, the charge 
density was assumed to be constant. This leaves polarized vortices outside 
the reach of the effective model. In terms of the Ginzburg-Landau model the 
difference between a vortex and a skyrmion is that a vortex is 
fully polarized while inside the skyrmion core there is some number of 
electrons with reversed spins. Such a depolarization helps to minimize the 
total Coulomb energy. It costs some Zeeman energy but it is not an obstacle 
when the Zeeman coupling is sufficiently weak.

  In this paper we  consider some topological properties of skyrmions and 
vortices in the framework of the Ginzburg-Landau model. We show that the Magnus
force acting on vortices and skyrmions is the same. We also consider 
statistical interactions of skyrmions. We will show that during a
clockwise exchange of two skyrmions the wave-function picks up 
the geometrical phase
\begin{equation}\label{00}
\Gamma=[S+\frac{1}{2(2n+1)}] 
\int_{t_{1}}^{t_{2}} \frac{d}{dt}Arg[\xi_{1}(t)-\xi_{2}(t)] \;\;, 
\end{equation}
where $S$ is the skyrmion's spin. This formula applies also to 
vortices when we put $S=S_{vortex}=\frac{1}{(2n+1)}$. Thus the
Ginzburg-Landau treatement shows that the statistical interaction
is much stronger for skyrmions than for vortices. It is 
proportional to the number of reversed spins.

\section{The Ginzburg-Landau model and topological defects}

We wish to investigate the statistical interaction of skyrmions 
in the quantum Hall effect. We will describe this system by the 
Zhang-Hansson-Kivelson model \cite{zhk} generalized by Kane and Lee 
\cite{kl} to describe unpolarized quantum Hall systems.

The Lagrangian of the model is
\begin{equation}\label{100}
L=i\phi^{\dagger}(\partial_{t}+ia_{0})\phi
 -\frac{1}{2m}\mid (\partial_{k}+i(a_{k}+eA_{k}))\phi \mid^{2}
 +\frac{1}{4\Theta}
  \varepsilon^{\mu\nu\sigma}a_{\mu}\partial_{\nu}a_{\sigma}
 -\frac{1}{2}\lambda [\phi^{\dagger}\phi(\vec{x})-\rho_{0}]^{2}
 +\gamma[\phi_{\downarrow}^{\star}\phi_{\downarrow}
        -\phi_{\uparrow}^{\star}\phi_{\uparrow}] 
 +\mu[\phi_{\downarrow}^{\star}\phi_{\downarrow}
     +\phi_{\uparrow}^{\star}\phi_{\uparrow}]\;\;.
\end{equation}
Greek indices denote $0,1,2$, while Latin indices take values $1,2$.
When they are repeated, summation is understood. We use the signature 
$(+,-,-)$. $\phi$ is a two-component complex scalar field, 
$\phi=(\phi_{\downarrow},\phi_{\uparrow})$. $a_{\mu}$ is a statistical 
gauge field while $A_{k}=-\frac{1}{2}B\varepsilon_{kl}x^{l}$ is a gauge 
potential of the external uniform magnetic field $B$ directed down 
the $z$-axis.  $e\rho_{0}$ is the positive background charge density, 
related to the external magnetic field by $eB=2\Theta\rho_{0}$. For the boson 
field to represent fermions the parameter $\Theta$ must take one of the values 
$\Theta=(2n+1)\pi$, where $n$ is a nonnegative integer. $m$ is the 
effective electronic mass and $\gamma$ is the effective Zeeman coupling. 
$\mu$ is a chemical potential, chosen
so that the ground state of the system (\ref{100}) is, up to a gauge
transformation, the solution 
$\phi_{\downarrow}=\sqrt{\rho_{0}},\phi_{\uparrow}=0$
with the statistical gauge field "screening" the external magnetic
field,  $a_{k}+eA_{k}=0$. The chemical potential has to be $\mu=-\gamma$. 
In this ground state the system is fully polarized. The Lagrangian 
$(\ref{100})$ should be supplemented by the Coulomb interaction term
\begin{equation}\label{120}
-\frac{1}{2}\int d^{2}x' [\phi^{\dagger}\phi(\vec{x})-\rho_{0}]
              \frac{e^{2}/\varepsilon}{\mid\vec{x}-\vec{x'}\mid}
              [\phi^{\dagger}\phi(\vec{x'})-\rho_{0}] \;\;,
\end{equation}
where $\varepsilon$ is a dielectric constant of the host material.

There are two types of relevant topological excitations in the model. One 
of them is simply a vortex or a fully polarized quasihole. This 
configuration is given by the Ansatz 
\begin{eqnarray}\label{130}
&& \phi_{\downarrow}=f_{v}(r)e^{-i\theta} \;\;, \nonumber\\
&& \phi_{\uparrow}=0 \;\;, \nonumber\\
&& a_{\theta}=\frac{eB}{2}r+a_{v}(r) \;\;,\nonumber\\
&& a_{0}=b_{v}(r) \;\;.
\end{eqnarray}
The modulus $f_{v}(r)$ has to interpolate between $f_{v}(0)=0$ and 
$f_{v}(\infty)=\sqrt{\rho_{0}}$. The asymptote of $a_{v}(r)$ at infinity
is $a_{v}(r)\approx \frac{1}{r}$. The vortex is by definition restricted to 
the lower spin component. This restriction is consistent with field 
equations of the model (\ref{100}) for any value of the Zeeman coupling 
$\gamma$. Such a solution indeed exists as is well known from the studies 
on vortices in the Ginzburg-Landau model of the fully polarized quantum Hall 
effect \cite{ezawa}. The total charge is proportional to the statistical 
magnetic flux and equals to $e/(2n+1)$. The distribution of the electronic 
density is such that it vanishes in the very center of the vortex and tends 
to $-e\rho_{0}$  ouside the vortex core. This is not the most favourable 
charge distribution from the point of view of both the Coulomb interaction 
and the quartic self-interaction term. While the total charge is fixed, 
the Coulomb interaction tends to minimize the integral of the charge density 
squared. In other words it tends to make the charge distribution as diluted 
as possible. When we restrict to the spin-down component, the deviation of 
the charge density from $-e\rho_{0}$ must equal to $e\rho_{0}$ at the 
center of the vortex. This is a topological obstruction. At the same time 
the total charge is quantized so there is not much freedom left to make 
the charge distribution more dilute. However, for sufficiently small Zeeman 
coupling the configuration may find a way to minimize its Coulomb energy 
by nucleation of the spin-up component. A more general configuration, with 
more freedom to minimize the Coulomb energy, is
\begin{eqnarray}\label{140}
&& \phi_{\downarrow}=f_{s}(r)e^{-i\theta}    \;\;,\nonumber\\
&& \phi_{\uparrow}=g(r)               \;\;,\nonumber\\
&& a_{0}=\frac{eB}{2}r+a_{s}(r) \;\;,\nonumber\\
&& a_{\theta}=b_{s}(r)             \;\;.
\end{eqnarray}
Now we admit a nonzero spin-up component. We want it to be nonzero
in the vortex core, $g(0)\neq 0$, and to cost a minimal gradient energy so
the phase of the spin-up component has to be constant. 
For a finite energy configuration $g(\infty)=0$. Excitation of the upper 
component costs some Zeeman energy so it is not energetically favourable 
except for small effective Zeeman couplings $\gamma$, where the gain in 
Coulomb energy can prevail the loss in Zeeman energy. Note that with the 
Ansatz (\ref{140}) the average spin points up in the middle of the soliton
while it points down outside the core. This configuration is a skyrmion
but in the language of the untruncated model (\ref{100}). As compared to a 
uniform background with charge density $\rho_{0}$, the $\uparrow$ component
adds a negative charge $Q_{\uparrow}$,
\begin{equation}\label{150}
 Q_{\uparrow}=-2\pi e\int_{0}^{\infty} rdr\; g^{2}(r) \;\;,
\end{equation}
while the $\downarrow$ component contributes a positive electronic charge 
deficit 
\begin{equation}\label{160}
 Q_{\downarrow}=-2\pi e\int_{0}^{\infty} rdr\;[f^{2}(r)-\rho_{0}] \;\;.
\end{equation}
The total charge is the same as for the fully polarized vortex;
the two contributions add up to 
$Q_{\uparrow}+Q_{\downarrow}=\frac{e}{(2n+1)}$.

  Thus there is an interplay of two factors, namely the Coulomb
and the Zeeman energy. For strong Zeeman coupling vortices are the relevant
quasiparticles. At weak Zeeman coupling vortices are still solutions of 
field equations but they have higher energy than skyrmions. In this limit
skyrmions become the relevant quasiparticles.

\section{Magnus force acting on skyrmions and vortices}

  We apply the adiabatic approximation to investigate
the dynamics of vortices and skyrmions. The only terms in the Lagrangian 
(\ref{100}), which can contribute to the terms in the effective 
mechanical Lagrangian which are linear in velocity, are
\begin{equation}\label{300}
L^{(1)}_{eff}=\int d^{2}x\; [ i\phi^{\dagger}\partial_{t}\phi
           -\frac{1}{4\Theta}\varepsilon^{kl}a_{k}\partial_{t}a_{l} ]\;\;. 
\end{equation}
The prescription for the adiabatic approximation is as follows.
Take the static vortex or skyrmion solution described by
the fields $\{ \phi(\vec{x}-\vec{\xi}), a_{\mu}(\vec{x}-\vec{\xi}) \}$
and located at an arbitrary position $\vec{\xi}$. Promote the parameter
$\vec{\xi}$ to the role of a time-dependent collective coordinate
$\vec{\xi}(t)$. In this way the static fields become time-dependent.
The final step is to substitute such a time-dependent field
configurations to the Lagrangian and integrate out their spatial
dependence. After the spatial integration one should be left with a purely 
mechanical Lagrangian being a functional of the trajectory $\vec{\xi}(t)$.

  The first term in (\ref{300}) does not contribute to the effective
Lagrangian. The static field is in the Coulomb gauge, $\partial_{k}a_{k}=0$,
so that the gauge field can be expressed as 
$a_{k}=\varepsilon_{kl}\partial_{l}U$, where $U$ is an auxillary
potential. In the adiabatic approximation we replace
$U(\vec{x})\rightarrow U[\vec{x}-\vec{\xi}(t)]$. The second term in
(\ref{300}) becomes
\begin{equation}\label{310}
\int d^{2}x\; 
[-\frac{1}{4\Theta}\varepsilon^{kl}\varepsilon_{km}\partial_{m}U
   \partial_{t}(\varepsilon_{ln}\partial_{n}U)]= 
\int d^{2}x\; 
[-\frac{1}{4\Theta}\varepsilon_{mn}\partial_{m}U\partial_{t}\partial_{n}U] 
\;\;.
\end{equation}
It can be shown, by an integration by parts, that this term does not
contribute to the effective Lagrangian. Thus the only contribution to the 
part of the effective mechanical Lagrangian which is linear in velocity comes 
from
\begin{equation}\label{320}
L^{(1)}_{eff}= \int d^{2}x \; i\phi^{\dagger}\partial_{t}\phi \;\;.
\end{equation}
When we introduce moduli and phases of the scalar fields,
$\phi_{A}=\sqrt{\rho_{A}}e^{i\chi_{A}}$ with $A=\uparrow,\downarrow$, the 
effective Lagrangian will split into contributions from spin-up and 
spin-down components 
\begin{equation}\label{330}
L^{(1)}_{eff}= -\int d^{2}x \;
[ \rho_{\uparrow}\partial_{t}\chi_{\uparrow}
 +\rho_{\downarrow}\partial_{t}\chi_{\downarrow}  ] 
\;\;. \end{equation}

  Let us apply the procedure to a single vortex or skyrmion.
The static configuration is described by (\ref{140}). In the vortex
case we can put formally $g(r)=0$. Let us take the trajectory
$\vec{\xi}(t)$ which passes through the origin at $t=0$. 
For a very small $\vec{\xi}$ the fields in (\ref{330}) can be expanded
as
\begin{eqnarray}\label{340}
&&\rho_{A}(\vec{x}-\vec{\xi})=\rho_{A}(r)
  -(\xi^{1}\cos\theta+\xi^{2}\sin\theta)\frac{d\rho_{A}}{dr}(r)
                                +O(\mid\xi\mid^{2})  \;\;,\nonumber\\
&&\partial_{t}\chi_{A}(\vec{x}-\vec{\xi})=
  -n_{A}(-\dot{\xi}^{1}\frac{\sin\theta}{r}
         +\dot{\xi}^{2}\frac{\cos\theta}{r})+O(\mid\xi\mid) \;\;.
\end{eqnarray}
With this expansion the effective Lagrangian becomes
\begin{equation}\label{350}
L^{(1)}_{eff}=-\pi\varepsilon_{kl}\xi^{k}\dot{\xi}^{l}
\sum_{A}n_{A}[\rho_{A}(\infty)-\rho_{A}(0)]+
O(\mid\xi\mid^{2}) \;\;.
\end{equation}
The system with just one quasiparticle is translationally invariant.
The term proportional to $\varepsilon_{kl}\xi^{k}\dot{\xi}^{l}$
is the most general term linear in velocity which is, up to a total
time derivative, translationally invariant. This proves that
the term  $O(\mid\xi\mid^{2})$ in Eq.(\ref{350}) vanishes identically.
 
  There are two contributions to the effective Lagrangian.
The contribution from the spin-up component vanishes
because $\rho_{\uparrow}(\infty)=n_{\uparrow}\rho_{\uparrow}(0)=0$. Thus
the only contribution comes from the lower component
$(\; \rho_{\downarrow}(0)=0\;,\;\rho_{\downarrow}(\infty)=\rho_{0} \;)$
and amounts to
\begin{equation}\label{360}
L^{(1)}_{eff}=\pi\rho_{0}\varepsilon_{kl}\xi^{k}\dot{\xi}^{l}\equiv
              \frac{eB}{2(2n+1)}\varepsilon_{kl}\xi^{k}\dot{\xi}^{l}. 
\end{equation}
The Magnus force is exactly the same for skyrmions and for vortices. 
It equals the Lorenz force acting on a particle with the charge 
$\frac{e}{2n+1} $. As discussed in \cite{stone} this is not a mere 
coincidence.
A quasiparticle is not an independent object but a defect composed out 
of electrons. As the location of defect moves it act with a Magnus force
on the surrounding electrons. They in turn respond with a current
which interacts through the Lorentz force with the external magnetic 
field. Thus the two forces combine to be just one.

   If the inertial mass were zero then the Magnus force would prevent a 
quasiparticle from moving with respect to the condesate. Note that in our 
derivation we have been slowly varying the location of a quasiparticle 
without moving the condesate so that the $\xi$'s in Eq.(\ref{360}) are 
quasiparticle coordinates with respect to the condesate's frame. 
The Lorentz force acts on the condensate as a whole. Namely, if the fields
$\{\; \phi(t,\vec{x}), a_{\mu}(t,\vec{x}) \;\}$ are solutions of field 
equations of the model (\ref{100}), then there are also the following
boosted solutions
\begin{eqnarray}\label{370}
&&\tilde{\phi}(t,\vec{x})= \phi[t,\vec{x}-\vec{R}(t)] e^{i\chi_{B}}
                                                       \;\;,\nonumber\\
&&\tilde{a}_{0}(t,\vec{x})= a_{0}[t,\vec{x}-\vec{R}(t)]
            -\dot{R}^{k}(t) a_{k}[t,\vec{x}-\vec{R}(t)]\;\;,\nonumber\\
&&\tilde{a}_{k}(t,\vec{x}) = a_{k}[t,\vec{x}-\vec{R}(t)]\;\;,\nonumber\\
&&\chi_{B}=m\dot{R}^{k}x^{k}-\frac{1}{2}m[\dot{R}^{k}\dot{R}^{k}]t
  -e\int_{t_{0}}^{t_{1}}d\tau\; \dot{R}^{k}(\tau)A_{k}[\vec{x}-\vec{R}(\tau)]
                                                       \;\;,\nonumber\\
&&\vec{R}(t_{0})=0                                       \;\;,\nonumber\\
&&\ddot{R}^{k}=-\frac{eB}{m}\varepsilon^{kl}\dot{R}^{l}  \;\;,\nonumber\\
&&A_{k}(\vec{x})=-\frac{B}{2}\varepsilon_{kl}x^{l}       \;\;.
\end{eqnarray}
Any solution can be boosted to move along an electronic cyclotron orbit.
Even the uniform condesate feels the Lorentz force in spite of the fact 
that the external magnetic field seems to be screened by the statistical 
gauge field. If there is a quasiparticle in the original solution then
after the boost it follows the cyclotron orbit but it does not move
with respect to the condensate. The Magnus force remains zero.

\section{Statistical interactions of quasiparticles}

  We have shown that the Magnus force is the same for both
skyrmions and vortices. Now we proceed to their mutual statistical
interactions. We will show that the statistical interaction
of skyrmions depends on the total number of inversed spins
they carry. 

Once again we will apply the formula (\ref{330}) but this time
to the system of two distant antiskyrmions. The total phase of the lower 
component reads 
\begin{equation}\label{500}
\chi_{\downarrow}(\vec{x})=-Arg(\vec{x}-\vec{\xi}_{1})
                         -Arg(\vec{x}-\vec{\xi}_{2})
\end{equation}
while the upper component phase is constant. Thus, once again, only the 
$\downarrow$ component contributes to the effective Lagrangian. The formula
(\ref{330}) simplifies to
\begin{equation}\label{510}
L^{(1)}_{eff}= -\int d^{2}x \;
[\delta\rho_{\downarrow}\partial_{t}\chi_{\downarrow}]\;\;, 
\end{equation}
where $\delta\rho_{\downarrow}=\rho_{\downarrow}-\rho_{0}$ is a deviation
of the down component's density from the uniform
background. When the distance between the skyrmions is large
as compared to their widths the deviation can be approximated by the
sum 
\begin{equation}\label{520}
\delta\rho_{\downarrow}(t,\vec{x})
                  \approx\delta\rho_{\downarrow}[\vec{x}-\vec{\xi}_{1}(t)]
                        +\delta\rho_{\downarrow}[\vec{x}-\vec{\xi}_{2}(t)] 
\;\; \end{equation}
of two nonoverlapping rotationally-symmetric deviations. When the
formulas (\ref{500}) and (\ref{520}) are substituted to (\ref{510}),
there appear four terms to be integrated out. Two of them
are "self-interaction" terms which lead to Magnus forces
as discussed in the previous section but in addition there are two
mutual-interaction terms
\begin{equation}\label{530}
L^{(1)}_{eff}=\int d^{2}x \;
\{ \delta\rho_{\downarrow}[\vec{x}-\vec{\xi}_{1}(t)]
   \partial_{t}Arg[\vec{x}-\vec{\xi}_{2}(t)]+
   \delta\rho_{\downarrow}[\vec{x}-\vec{\xi}_{2}(t)]
   \partial_{t}Arg[\vec{x}-\vec{\xi}_{1}(t)] \} \;\;.
\end{equation}
For very distant skyrmions the density deviations can be 
approximated by 
$ \delta\rho_{\downarrow}[\vec{x}-\vec{\xi}_{1}(t)]\approx 
  -\frac{Q_{\downarrow}}{e}\delta^{(2)}[\vec{x}-\vec{\xi}_{1}(t)]$,
where $Q_{\downarrow}$ is the charge deficit of the spin-down component,
compare with (\ref{160}). In this approximation the integral (\ref{530}) 
becomes
\begin{equation}\label{535}
L^{(1)}_{eff}= -\frac{Q_{\downarrow}}{e}\int d^{2}x \;
\{ \delta^{(2)}[\vec{x}-\vec{\xi}_{1}(t)]
   \partial_{t}Arg[\vec{x}-\vec{\xi}_{2}(t)]+
   \delta^{(2)}[\vec{x}-\vec{\xi}_{2}(t)]
   \partial_{t}Arg[\vec{x}-\vec{\xi}_{1}(t)] \} \equiv
   -\frac{Q_{\downarrow}}{e}\frac{d}{dt}Arg[\xi_{1}(t)-\xi_{2}(t)]\;\;,
\end{equation}
which is just the statistical interaction we were looking for.
The prefactor $Q_{\downarrow}/e$ is the total number of electrons missing
in the lower component as compared to the uniform ground state. As the 
number of electrons in the upper component is $-Q_{\uparrow}/e$, the 
total spin of the skyrmion as compared to the uniform background is 
$S=\frac{1}{2}(Q_{\uparrow}-Q_{\downarrow})/e$.  
$Q_{\uparrow}=-Q_{\downarrow}-\frac{e}{2n+1}$, so that the statistical 
interaction becomes
\begin{equation}\label{540}
L^{(1)}_{eff}=-[S+\frac{1}{2(2n+1)}] \frac{d}{dt}Arg[\xi_{1}(t)-\xi_{2}(t)] 
\;\;, 
\end{equation}
where $S$ is the total spin of the skyrmion with respect to the uniform
polarized background. In the case of the vortex $S=S_{vortex}=1/2(2n+1)$.

\section{Dual formulation of the model}

  To get further insight into statistical interactions of vortices
we will perform Hubbard-Stratanovich transformation on the model
(\ref{100}). The phase gradients squared in the Lagrangian (\ref{100})
can be rewritten as  
\begin{equation}\label{700}
\rho_{A}^{\frac{3}{2}}\exp i\int d^{2}x\;
    [-\frac{\rho_{A}}{2m}\partial_{k}\chi_{A}\partial_{k}\chi_{A}]=
\int [DI_{k}^{A}]\exp i\int d^{2}x\;
    [\frac{I^{A}_{k}I^{A}_{k}}{2m\rho_{A}}-
     \frac{I^{A}_{k}\partial_{k}\chi_{A}}{m}] \;\;,
\end{equation}
where the auxillary fields $I^{A}_{k}$ have been introduced.
In the next step the phases can be split into two parts
$\chi_{A}=\chi^{0}_{A}+\eta_{A}$. $\chi^{0}$'s are multivalued
phases due to vortices/skyrmions while $\eta$'s are singlevalued
components. At the present stage the Lagrangian is linearized
in phase gradients. Functional integration over $\eta$'s leads to the 
conservation laws
\begin{equation}\label{710}
\dot{\rho}_{A}
+\partial_{k}[\frac{I^{A}_{k}+\rho_{A}(a_{k}^{A}+eA_{k})}{m}]=0 \;\;.
\end{equation}
These conservation laws will be identically satisfied when we
introduce a pair of dual gauge fields
\begin{equation}\label{720}
\{\; \rho_{A} \;,\; \frac{1}{m}[I^{A}_{k}+\rho_{A}(a_{k}^{A}+eA_{k})] \;\}=
\varepsilon^{\mu\nu\sigma}\partial_{\nu}B^{A}_{\sigma}\stackrel{def}{=}
\frac{1}{2}\varepsilon^{\mu\nu\sigma}H^{A}_{\nu\sigma} \;\;.
\end{equation}
This definition together with the definition of the topological
currents, $2\pi K^{\mu}_{A}=
\varepsilon^{\mu\nu\sigma}\partial_{\nu}\partial_{\sigma}\chi_{A}$,
leads, after some rearrangement and integration by parts, to the
following dual Lagrangian
\begin{equation}\label{730}
L_{D}=L_{B}+
\frac{1}{4\Theta}\varepsilon^{\mu\nu\sigma}a_{\mu}\partial_{\nu}a_{\sigma}-
\sum_{A}[ \varepsilon^{\mu\nu\sigma}(a_{\mu}+eA_{\mu})
                                              \partial_{\nu}B^{A}_{\sigma}+
2\pi B_{\mu}^{A}K^{\mu}_{A} ] \;\;.
\end{equation}
$L_{B}$ is the Lagrangian of the dual gauge field
\begin{equation}\label{740}
L_{B}=
\sum_{A}\{  \frac{m H^{A}_{ok} H^{A}_{ok}}{2 H^{A}_{12}}
-\frac{ \partial_{k}H^{A}_{12} \partial_{k}H^{A}_{12}}{8m H^{A}_{12}} \}
-\frac{\lambda}{2}[(\sum_{A} H^{A}_{12}) -\rho_{0}]^{2}-2\gamma 
H^{\uparrow}_{12} \end{equation}
Let us concentrate now on the minimal coupling between the topological
current and the dual gauge field. The topological current due
to the antivortex/antiskyrmion moving along the trajectory $\vec{\xi}(t)$ is
\begin{equation}\label{750}
K^{0}_{A}=-\delta_{\downarrow A}\delta^{2}[\vec{x}-\vec{\xi}(t)] \;\;,\;\;
K^{k}_{A}=-\delta_{\downarrow A}
\dot{\xi}^{k}(t)\delta^{2}[\vec{x}-\vec{\xi}(t)] \;\;. 
\end{equation}
Only the topological current of the $\downarrow$ component is nonzero.
It is like a current of a negatively charged point particle.
What is the dual gauge potential which couples
to the topological current? We know that the relation between the dual 
magnetic field and the original density is $\rho_{A}=-H^{A}_{21}$.
Far from any topological defects the dual magnetic field is
just $H^{A}_{21}=-\delta_{\downarrow A}\rho_{0}$ and is directed down
the $z$-axis. A single vortex/skyrmion moves in a uniform dual magnetic 
field. This is the origin of the Magnus force. Thus Magnus force is
in fact a dual Lorentz force. Topological defects distort
the uniform background. Their contribution to the dual gauge potential 
is 
\begin{equation}\label{760}
B^{\downarrow}_{k}(\vec{x})=
\frac{Q_{\downarrow}}{2\pi e}\sum_{v}\varepsilon_{kl}
\frac{x^{l}-\xi^{l}_{v}}{\mid \vec{x}-\vec{\xi}_{v} \mid^{2}} \;\;,
\end{equation}
where $v$ runs over topological defects.
With this form of the gauge potential we are able to work out
the mutual statistical interaction
\begin{equation}\label{770}
-2\pi\int d^{2}x\;K^{\mu}_{A}B_{\mu}^{A}=
\frac{Q_{\downarrow}}{e}\sum_{v\neq w}\int d^{2}x\;
\varepsilon_{kl}
\frac{x^{l}-\xi^{l}_{v}}{\mid \vec{x}-\vec{\xi}_{v} \mid^{2}} 
\dot{\xi}^{k}_{w}\delta^{(2)}(\vec{x}-\vec{\xi}_{w})=
-\frac{Q_{\downarrow}}{e}\sum_{v<w}\frac{d}{dt}Arg[\vec{\xi}_{v}-\vec{\xi}_{w}]
\end{equation}
which is the same as Eq.(\ref{540}).

  Even if the inertial mass of quasiparticles is zero there still remains
some possibility of motion provided that there are long range potential 
interactions between them. For nonzero Zeeman coupling the quasiparticles
are exponentially localized field configurations. There is thus no long-range
potential interaction through the matter fields. However there is still
the long range Coulomb interaction (\ref{120}) which gives rise
to the following effective Lagrangian for diluted quasiholes
\begin{equation}\label{780}
L^{(1)}_{eff}=\pi\rho_{0}\sum_{v}\varepsilon_{kl}\xi_{v}^{k}\dot{\xi}_{v}^{l}
-\frac{e^{2}}{\varepsilon(2n+1)^{2}}\sum_{v<w}
\frac{1}{\mid\vec{\xi}_{v}-\vec{\xi}_{w}\mid}
-\sum_{v<w}[S+\frac{1}{2(2n+1)}]\frac{d}{dt}Arg[\vec{\xi}_{v}-\vec{\xi}_{w}]
\end{equation}
The last term is a total time derivative and can be neglected in 
classical considerations. According to the classical part of this Lagrangian
the particles move along the trajectories of constant total potential
energy. For example a pair of quasiholes performs a strictly circular
clockwise motion.

\section{Remarks}

  We have derived statistical interaction of skyrmions in the framework
of the time-dependent Ginzburg-Landau model. These interactions can be much
stronger than interactions of fully polarized vortices. This conclusion
could not be obtained in the effective Heisenberg model. This example
shows that the Heisenberg model is not relevant not only to description
of quasiparticles' energies but it also fails to give a correct answer
to more basic topological questions. 

  We did not make any quantitative predictions as to what is the 
skyrmion's spin $S$ which determines the strenght of the statistical
interaction. Such calculations are possible in the Ginzburg-Landau model 
but, there exist already microscopic Hartree-Fock 
techniques to establish the value of $S$, see \cite{fertig}. 

  From our calculations it turns out that that the geometrical
phase picked up by electronic wave-function during a clockwise exchange
of two skyrmions is 
\begin{equation}\label{ff1}
\Gamma = \pi [S+\frac{1}{2(2n+1)}] \equiv
\pi [N_{\uparrow}+\frac{1}{(2n+1)}] \;\;,
\end{equation}
where $N_{\uparrow}$ is the number of electrons with reversed
spins trapped inside the skyrmion core. This phase determines the 
quantum statistics of skyrmions. There are two types of anyons.
For $N_{\uparrow}$ even the phase is $\frac{\pi}{(2n+1)}$ up to 
an even multipilicity of $\pi$. This is the same kind of statistics
as for fully polarized Laughlin quasiparticles (or quasiholes). Condensation
of such skyrmions leads to hierarchy states with odd denominator
filling fractions. For skyrmions with $N_{\uparrow}$ odd the exchange
phase (\ref{ff1}) is $\pi\frac{2n+2}{2n+1}$ up to unrelevant even 
multiplicities of $\pi$. This type of anyons differs from the Lauglin
quasiparticles by an addition of one flux quantum. Their condensation
gives rise to even-denominator filling fractions. To summarize
the condesation of skyrmions of the primary state with 
$\nu=\frac{1}{2n+1}$ gives rise to the following filling fractions
labelled by p
\begin{equation}\label{ff2}
\nu=\frac{1}{(2n+1)-\frac{\alpha}{p}} \;\;,\;\; p=1,2,3.... \;\;,
\end{equation}
where $\alpha=+1$ or $-1$ dependent on whether skyrmions or antiskyrmions
condense. Note that in the above formula $p$ is an arbitrary integer
so that the filling fractions can have both even and odd denominators.

\paragraph{Aknowledgement.}

I would like to thank Wojtek Zakrzewski and Jens Gladikowski for useful
discussions and inspiring atmosphere. This research was supported
in part by the KBN grant 2P03B08508 and in part by Foundation for Polish
Science.


\begin{thebibliography}{99}

\bibitem{sondhi} S.L.Sondhi, A.Karlshede, S.A.Kivelson and E.H.Rezayi,
                                           Phys.Rev.B 47, 16419 (1993),
\bibitem{fertig} H.A.Fertig, L.Brey, R.Cote and A.H.MacDonald,
                                           Phys.Rev.B 50, 11018 (1994),
\bibitem{moon} K.Moon, H.Mori, Kun Yang, S.M.Girvin, A.H.MacDonald,
               L.Zheng, D.Yoshioka and S.-C.Zhang, Phys.Rev.B 51, 5138 (1995),
\bibitem{knight} S.E.Barrett, G.Dabbagh, L.N.Pfeiferer, K.W.West, R.Tycko,
                 Phys.Rev.Lett. 74, 5112 (1995), R.Tycko, S.E.Barrett,
                 G.Dabbagh, L.N.Pfeiferer, K.W.West, Science 268, 1460 (1995)
\bibitem{stone} M.Stone, preprint cond-mat/9512010,

\bibitem{zhk} S.-C.Zhang, T.H.Hansson, S.A.Kivelson, Phys.Rev.Lett. 
                                                                62 (1989) 82,
\bibitem{kl} D.H.Lee and C.L.Kane, Phys.Rev.Lett. 64, 1313 (1990)

\bibitem{ezawa} Z.F.Ezawa, M.Hotta and A.Iwazaki, Phys.Rev.B 46, 7765 (1992),



\end{thebibliography}
\end{document}